\documentstyle [12pt] {article}

\catcode`@=12
\input epsf.tex
\def\DESepsf(#1 width #2){\epsfxsize=#2 \epsfbox{#1}}

\begin{document}
\thispagestyle{empty}
\begin{flushright} UCRHEP-T196\\Fermilab Pub-97/262-T\\
July 1997\
\end{flushright}
\vspace{0.3in}
\begin{center}
{\Large	\bf Lower Bound on the Pseudoscalar Mass in\\}
\vspace{0.1in}
{\Large \bf the Minimal Supersymmetric Standard Model\\}
\vspace{1.0in}
{\bf E. Keith$^1$, Ernest Ma$^1$, and D. P. Roy$^{2,3}$\\}
\vspace{0.3in}
{$^1$ \sl Department of Physics, University of California\\}
{\sl Riverside, California 92521, USA\\}
\vspace{0.1in}
{$^2$ \sl Tata Institute of Fundamental Research, Mumbai 400 005, India\\}
\vspace{0.1in}
{$^3$ \sl Fermi National Accelerator Laboratory, Batavia, Illinois 60510, USA\\}
\vspace{0.5in}
\end{center}
\begin{abstract}\
In the Higgs sector of the Minimal Supersymmetric Standard Model, the mass 
of the pseudoscalar $A$ is an independent parameter together with $\tan \beta 
\equiv v_2/v_1$.  If $m_A$ is small, then the process $e^+ e^- \rightarrow 
h + A$ is kinematically allowed and is suppressed only if $\tan \beta$ is 
small.  On the other hand, the mass of the charged Higgs boson is now near 
$M_W$, and the decay $t \rightarrow b + h^+$ is enhanced if $\tan \beta$ is 
small.  Since the former has not been observed, and the branching fraction 
of $t \rightarrow b + W$ cannot be too small (by comparing the experimentally 
derived $t \bar t$ cross section from the leptonic channels 
with the theoretical prediction), we can infer a phenomenological 
lower bound on $m_A$ of at least 60 GeV for all values of $\tan \beta$.
\end{abstract}

\newpage
\baselineskip 24pt

The most studied extension of the standard $SU(2) \times U(1)$ electroweak 
gauge model is that of supersymmetry with the smallest necessary particle 
content.  In this Minimal Supersymmetric Standard Model (MSSM), there are 
two scalar doublets $\Phi_1 = (\phi_1^+, \phi_1^0)$ and $\Phi_2 = (\phi_2^+, 
\phi_2^0)$, with Yukawa interactions $\overline {(u,d)}_L d_R \Phi_1$ and 
$\overline {(u,d)}_L u_R \tilde \Phi_2$, respectively, where $\tilde \Phi_2 = 
i \sigma_2 \Phi_2^* = (\overline {\phi_2^0}, -\phi_2^-)$.  The Higgs sector 
of the MSSM has been studied in great detail\cite{1} and it is a current 
topic of intensive experimental and theoretical scrutiny.\cite{2}  There 
are five physical Higgs bosons in the MSSM:  two neutral scalars 
($h$ and $H$), one neutral pseudoscalar ($A$), and two charged ones 
($h^\pm$).  Their masses and couplings to other particles are completely 
determined up to two unknown parameters which are often taken to be 
$m_A$ and $\tan \beta \equiv v_2/v_1$, where $v_i$ is the vacuum expectation 
value of $\phi_i^0$.

In the following, we will show that $m_A > 60$ GeV for all values of 
$\tan \beta$.  Our conclusion is based on a combination of theoretical 
and experimental inputs from a number of different observations which 
have become available recently.

In the MSSM, the pseudoscalar Higgs boson $A$ and the charged Higgs bosons 
$h^\pm$ are given by analogous expressions, namely
\begin{eqnarray}
A &=& \sqrt 2 (\sin \beta {\rm Im} \phi_1^0 - \cos \beta {\rm Im} \phi_2^0), 
\\ h^\pm &=& \sin \beta \phi_1^\pm - \cos \beta \phi_2^\pm.
\end{eqnarray}
At tree level, their masses are related by $m_{h^\pm}^2 = m_A^2 + M_W^2$. 
The mass-squared matrix spanning the two neutral scalar Higgs bosons 
$\sqrt 2 {\rm Re} \phi^0_{1,2}$ is given by
\begin{equation}
{\cal M}^2 = \left[ \begin{array} {c@{\quad}c} m_A^2 \sin^2 \beta + M_Z^2 
\cos^2 \beta & -(m_A^2 + M_Z^2) \sin \beta \cos \beta \\ -(m_A^2 + M_Z^2) 
\sin \beta \cos \beta & m_A^2 \cos^2 \beta + M_Z^2 \sin^2 \beta + 
\epsilon/\sin^2 \beta \end{array} \right].
\end{equation}
In the above, $\epsilon$ is the leading radiative correction\cite{6} due to the 
$t$ quark:
\begin{equation}
\epsilon = {{3 g_2^2 m_t^4} \over {8 \pi^2 M_W^2}} \ln \left( 1 + 
{\tilde m^2 \over m_t^2} \right),
\end{equation}
where $\tilde m$ is the mass parameter for the supersymmetric scalar quarks.

Let us take $m_A = 0$ and rotate ${\cal M}^2$ to the basis spanned by 
\begin{equation}
h_1 = \sqrt 2 (\sin \beta {\rm Re} \phi_1^0 - \cos \beta {\rm Re} \phi_2^0), 
~~~ h_2 = \sqrt 2 (\cos \beta {\rm Re} \phi_1^0 + \sin \beta {\rm Re} 
\phi_2^0).
\end{equation}
We get\cite{3}
\begin{equation}
{\cal M}^2 = \left[ \begin{array} {c@{\quad}c} M_Z^2 \sin^2 2 \beta + \epsilon 
\cot^2 \beta & -M_Z^2 \sin 2 \beta \cos 2 \beta + \epsilon \cot \beta \\ 
-M_Z^2 \sin 2 \beta \cos 2 \beta + \epsilon \cot \beta & M_Z^2 \cos^2 2 \beta 
+ \epsilon \end{array} \right].
\end{equation}
It is well-known that in this basis, the $h_1 Z Z$ and $h_2 A Z$ couplings 
are absent, hence the nonobservation of $e^+ e^- \rightarrow h + A$ does 
not rule out any value of $m_A$ if $\tan \beta$ is small enough\cite{4}.  
In this limit, 
the eigenstates of ${\cal M}^2$ are essentially $h_1$ and $h_2$.  If 
$h \simeq h_1$, then it is too heavy to be produced.  If $h \simeq h_2$, 
then its coupling to $A$ is too small to have a measurable branching 
fraction.  Note that $\epsilon \simeq M_Z^2$, {\it i.e.} (91 GeV)$^2$, 
for $m_t = 175$ GeV and $\tilde m = 1$ TeV.

From the nonobservation of $e^+ e^- \rightarrow h + Z$ where the $Z$ 
boson may be either real or virtual and the nonobservation of $e^+ e^- 
\rightarrow h + A$, where $h$ is an arbitrary linear combination of $h_1$ 
and $h_2$, it is possible to obtain the MSSM exclusion region in the 
$m_A - \tan \beta$ plane.  One such detailed analysis\cite{5} using only 
LEP1 data collected at the $Z$ resonance shows that $m_A$ has to be greater 
than about $M_Z/2$ for $\tan \beta > 1$.  With the higher energies 
available at LEP2 since then, this bound is expected to be at least 
60 GeV.

To obtain a lower bound on $m_A$ for $\tan \beta < 1$, we propose to use 
the MSSM relationship\cite{6}
\begin{equation}
m_{h^\pm}^2 = m_A^2 + M_W^2 - {\epsilon \over {4 \sin^2 \beta}} {M_W^2 
\over m_t^2},
\end{equation}
where the last term is the leading radiative correction for $\tan \beta < 1$. 
We then derive bounds on $m_A$ from the bounds on $m_{h^\pm}$ by considering 
$t$ decay.  Taking $m_t = 175$ GeV, we see that $t \rightarrow b + h^+$ is 
allowed for values of $m_{h^\pm}$ up to 170 GeV, corresponding to $m_A$ up 
to about 150 GeV.  The nonobservation of 
the above process would then translate into lower bounds on $m_A$ as a 
function of $\tan \beta$.

In the MSSM, the charged-Higgs-boson couplings to the quarks and leptons 
are given by
\begin{equation}
{\cal H}_{int} = {{-g_2} \over {\sqrt 2 M_W}} h^+ [ \cot \beta ~m_{u_i} 
\bar u_i d_{iL} + \tan \beta ~m_{d_i} \bar u_i d_{iR} + \tan \beta ~m_{l_i} 
\bar \nu_i l_{iR}] + h.c.,
\end{equation}
where the subscript $i$ represents the generation index, and we have 
used the diagonal KM matrix approximation\cite{7}.  The leading-logarithm 
QCD (quantum chromodynamics) correction is taken into account by 
substituting the quark mass parameters by their running masses evaluated 
at the $h^\pm$ mass scale.  The resulting decay widths are
\begin{equation}
\Gamma (t \rightarrow b h^+) = {{g_2^2 \lambda^{1/2} (1, m_b^2/m_t^2, 
m_{h^+}^2/m_t^2)} \over {64 \pi M_W^2 m_t}} [(m_t^2 \cot^2 \beta + 
m_b^2 \tan^2 \beta)(m_t^2 + m_b^2 - m_{h^+}^2) - 4 m_t^2 m_b^2],
\end{equation}
where $\lambda$ denotes the usual Kallen function and $\lambda^{1/2}$ 
is equal to the magnitude of the momentum of either decay product divided 
by $m_t/2$, and
\begin{eqnarray}
\Gamma (h^+ \rightarrow \tau^+ \nu) &=& {{g_2^2 m_{h^+}} \over {32 \pi 
M_W^2}} m_\tau^2 \tan^2 \beta, \\ 
\Gamma (h^+ \rightarrow c \bar s) &=& {{3 g_2^2 m_{h^+}} \over {32 \pi 
M_W^2}} (m_c^2 \cot^2 \beta + m_s^2 \tan^2 \beta).
\end{eqnarray}
Assuming that the only other competing channel is the standard-model decay 
$t \rightarrow b W^+$, the $t \rightarrow b h^+$ branching fraction is then
\begin{equation}
B = {{\Gamma (t \rightarrow b h^+)} \over {\Gamma (t \rightarrow b h^+) 
+ \Gamma (t \rightarrow b W^+)}},
\end{equation}
where
\begin{equation}
\Gamma (t \rightarrow b W^+) = {{g_2^2 \lambda^{1/2} (1, m_b^2/m_t^2, 
M_W^2/m_t^2)} \over {64 \pi M_W^2 m_t}} [M_W^2 (m_t^2 + m_b^2) + 
(m_t^2 - m_b^2)^2 - 2 M_W^4].
\end{equation}
It is clear from Eq.~(9) that $B$ has a minimum at $\tan \beta = 
(m_t/m_b)^{1/2} \simeq 6$, but it becomes large for $\tan \beta < 1$ and 
$\tan \beta > m_t/m_b$.  Thus we expect to see a sizeable $t \rightarrow 
b h^+$ signal in these two regions if $m_{h^+} < m_t$.

We see from Eqs.~(10) and (11) that $\tau^+ \nu$ is the dominant decay 
mode of $h^+$ if $\tan \beta >> 1$.  Thus an excess of $t \bar t$ events 
in the $\tau$ channel compared to the standard-model prediction constitutes 
a viable $h^\pm$ signal in the large $\tan \beta$ region.  A recent 
analysis\cite{8} of the CDF $t \bar t$ data in the $\tau l$ channel 
($l = e, \mu$) has led to a mass bound of $m_{h^\pm} > 100$ GeV for 
$\tan \beta > 40$.  A similar bound has also been obtained from the same 
$t \bar t$ data in the inclusive $\tau$ channel\cite{9}.

The above method is not applicable in the small $\tan \beta$ region, where 
$h^+$ is expected to decay mainly into $c \bar s$, {\it i.e.} two 
jets. On the other hand, we can use the so-called disappearance method to look 
for the presence of $t \rightarrow b h^+$ decay in both the small and large 
$\tan \beta$ regions\cite{7} as described below.  The key observation is 
that $h^\pm$ couples negligibly to the light fermions, particularly $e$ and 
$\mu$, whereas the $W$ boson couples to them with full strength universally. 
Since the $e$ and $\mu$ decay modes play an important role in the detection 
of $t \bar t$ events at the Tevatron, the experimentally derived $t \bar t$ 
cross section is sensitive to the branching fraction $B$ of Eq.~(12). After 
all, if $t$ decays into $b h^+$, there would not be any energetic $e$ or 
$\mu$ in the final state, as would be possible with the $W$ boson.

The experimental $t \bar t$ cross sections obtained by the CDF and D0 
collaborations\cite{10,11} are weighted averages of their measured cross 
sections in the (I) dilepton ($ll$) and (II) lepton plus multijet ($lj$) 
channels, using the standard formula
\begin{equation}
\sigma = {{\Sigma (\sigma_i / \delta_i^2)} \over {\Sigma (1/\delta_i^2)}}.
\end{equation}
They are summarized below.
\begin{equation}
{\rm CDF}: ~~\sigma_{ll} = 8.5 \begin{array} {c} +4.4 \\ -3.4 \end{array} 
{\rm pb}, ~~~ \sigma_{lj} = 7.2 \begin{array} {c} +2.1 \\ -1.7 \end{array} 
{\rm pb} ~~~ \Rightarrow ~~\sigma_{\rm CDF} = 7.5 \begin{array} {c} +1.9 \\ 
-1.6 \end{array} {\rm pb}.
\end{equation}
\begin{equation}
{\rm D0}: ~~ \sigma_{ll} = 6.3 \pm 3.3 {\rm ~pb}, ~~~ \sigma_{lj} = 5.1 \pm 
1.9 {\rm ~pb} ~~~ \Rightarrow ~~\sigma_{\rm D0} = 5.5 \pm 1.8 {\rm ~pb}.
\end{equation}
The $\sigma_{lj}$ of CDF is a weighted average of the measured cross 
sections using the SVX and SLT $b$-tagging methods; that of D0 is a 
weighted average of those using kinematic cuts and SLT $b$-tagging.  
In both cases, the weight of the SLT method is 
rather low.  From Eqs.~(15) and (16), we see that for both CDF and D0, 
$\delta_{lj} \simeq \delta_{ll}/2$, hence
\begin{equation}
\sigma \simeq {{\sigma_{ll} + 4 \sigma_{lj}} \over 5}.
\end{equation}
Furthermore, since the CDF and D0 cross sections have essentially identical 
errors, we can take a simple average of the two:
\begin{equation}
\sigma_{\rm CDF + D0} = 6.5 \begin{array} {c} +1.3 \\ -1.2 \end{array} 
{\rm pb}.
\end{equation}
Here we have combined the two errors using $\delta^{-2} = \delta_1^{-2} + 
\delta_2^{-2}$, since they are largely statistical.

We note that the dilepton channel (I) corresponds to the leptonic ($e, \mu$) 
decay of both the $t$ and $\bar t$ quarks, whereas the lepton plus multijet 
channel (II) corresponds to the leptonic decay of one, say $t \rightarrow b 
l^+ \nu$, and the hadronic decay of the other.  For the standard-model decay 
$t \rightarrow b W^+$, the respective branching fractions are 2/9 and 2/3, 
whereas for the postulated decay $t \rightarrow b h^+$, they are 0 and 
a function which rises rapidly to 1 for $\tan \beta < 1$.  Thus the relative 
contributions of different final states to the two channels are $W W : 
W h^\pm : h^\pm h^\mp = 1 : 0 : 0$ for ($ll$) and $1 : 3/4 : 0$ for ($lj$).  
[We have used the maximum value of 3/4 corresponding to very small $\tan 
\beta$.  This is a conservative approach, because any smaller value will 
give us a better bound on $m_{h^\pm}$ as explained below.]  We have then 
a suppression factor relative to the standard model of
\begin{equation}
f_{ll} = (1 - B)^2 \simeq 0.5 ~~({\rm for} ~B = 0.3),
\end{equation}
\begin{equation}
f_{lj} = (1-B)^2 + 2B(1-B)(3/4) \simeq 0.8 ~~({\rm for} ~B = 0.3).
\end{equation}
Since the relative weights of the ($ll$) and ($lj$) channels are 1:4, 
Eqs.~(19) and (20) correspond to an effective suppression 
factor of
\begin{equation}
f = 0.74 ~~({\rm for} ~B = 0.3).
\end{equation}
We note that for large $\tan \beta$, $h^\pm$ decays mainly into $\tau$, 
hence it would be hard for the $W h^\pm$ final state to pass the 
$n_{\rm jet} \geq 3$ cut required for the ($lj$) channel.  This implies 
an extra suppression factor of about 1/3 for the $W h^\pm$ contribution, 
hence $f$ is about 0.7 already for $B = 0.2$, {\it i.e.} our bound is 
conservative because it assumes $B = 0.3$.

Finally the theoretical estimates of the $t \bar t$ cross section including 
higher-order QCD corrections are 4.13 to 5.48 pb\cite{12}, and 
5.10 to 5.59 pb\cite{13}.  These ranges are not identical, but the two 
estimates are in reasonable agreement as to their upper bounds.  We shall thus 
assume for our purpose that
\begin{equation}
\sigma (t \bar t) \leq 5.6 ~{\rm pb}.
\end{equation}
Combining this with the suppression factor of Eq.~(21), we obtain an upper 
bound of
\begin{equation}
\sigma \leq 4.1 ~{\rm pb}
\end{equation}
for the weighted cross section of Eq.~(17).  This is $2\sigma$ lower than 
the combined CDF and D0 estimate of Eq.~(18), as well as the CDF estimate 
of Eq.~(15).  Hence we can take $B = 0.3$ as a $2\sigma$ upper bound for the 
branching fraction of $t \rightarrow b h^+$ decay.  In Figure 1 we plot the 
exclusion regions of $m_{h^\pm}$ as a function of $\tan \beta$ using $B = 
0.3$.  We also show the exclusion region obtained in Ref.~[8], which used the 
``appearance" method of looking for $\tau$, instead of the ``disappearance" 
method of not finding $e$ or $\mu$ discussed here.

To convert a bound on $m_{h^\pm}$ to one on $m_A$, we use the full 
expression including all one-loop radiative corrections\cite{6} in place 
of Eq.~(7) which is approximate and valid only for $\tan \beta < 1$.  
In Figure 2 we plot 
the exclusion regions of $m_A$ as a function of $\tan \beta$ deduced from 
$t$ decay and $t \bar t$ production corresponding to Fig.~1.  We note that 
the radiative correction is negative for small $\tan \beta$ which 
increases the $m_A$ bound, and is positive for large $\tan \beta$ which 
decreases it.  We note also that at extreme values of $\tan \beta$, near 0.2 
and 100, the Yukawa couplings involved are becoming too large for a 
perturbative calculation to be reliable.  We then add a line 
at $m_A = 60$ GeV for $\tan \beta > 1$ as a conservative upper limit 
from the combined LEP data\cite{5,14}.   Our conclusion is simple: in the 
Minimal Supersymmetric Standard Model, combining what we know from LEP and the 
Tevatron and using a conservative estimate of the theoretical $t \bar t$ 
cross section, the pseudoscalar mass $m_A$ is now known to be greater 
than 60 GeV for all values of $\tan \beta$.
\vspace{0.3in}

\noindent {\bf Note Added}: After the completion of our paper, we found out 
that the ALEPH Collaboration has just recently obtained\cite{15} the bound 
$m_A > 62.5$ GeV for $\tan \beta > 1$.
\vspace{0.3in}
\begin{center} {ACKNOWLEDGEMENT}
\end{center}

One of us (DPR) thanks N. K. Mondal of the D0 Collaboration for discussions.
This work was supported in part by the U. S. Department of Energy under 
Grant No. DE-FG03-94ER40837.
\vspace{0.3in}

\bibliographystyle{unsrt}

\begin{thebibliography}{99}
\bibitem{14} A detailed updated analysis (A. Sopczak, private communications) 
shows that the $m_A$ bound is actually much higher than 60 GeV.
\end{thebibliography}

\newpage
\begin{center} {\large \bf Figure Captions}
\end{center}

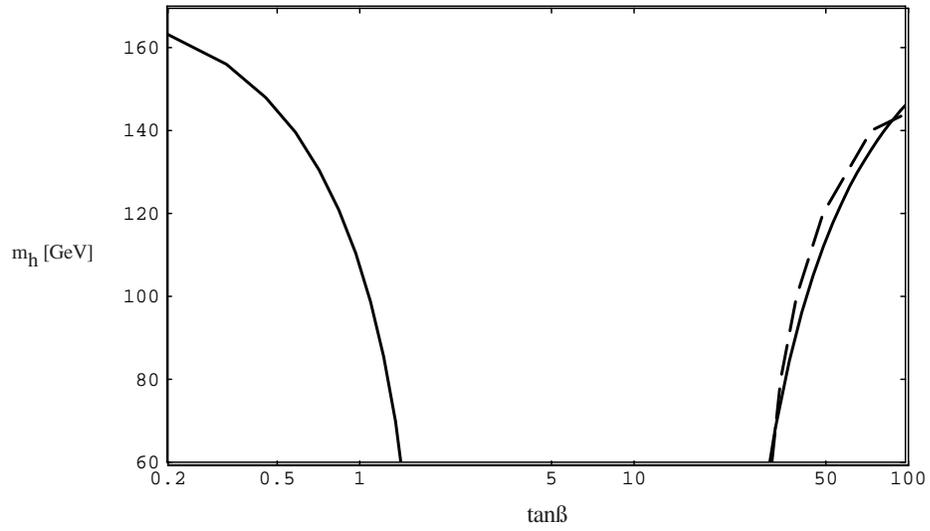
\noindent Fig.~1. Exclusion regions at 95\% confidence level in the 
$m_{h^\pm} - \tan \beta$ plane using $B = 0.3$ (solid lines) for $t 
\rightarrow b h^+$ as explained in the text.  The dashed line corresponds 
to the method used in Ref.~[8].

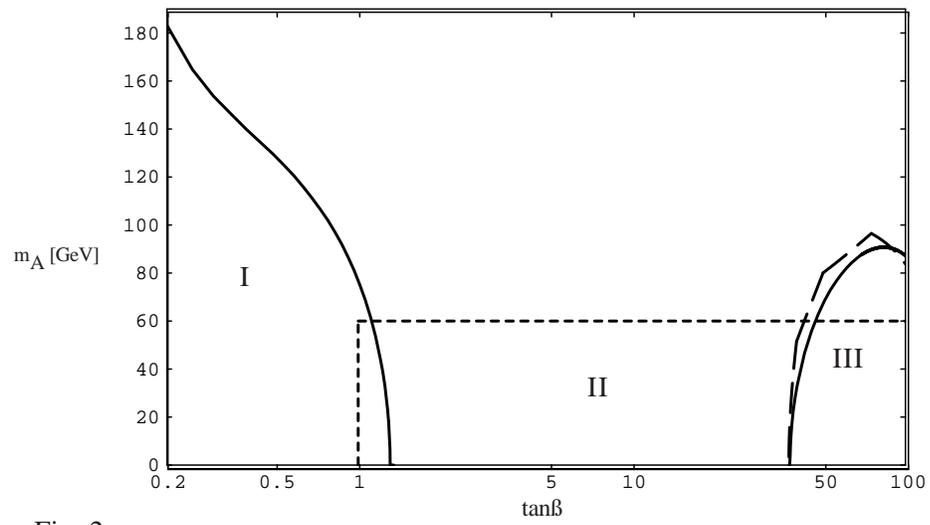
\noindent Fig.~2. Exclusion regions at 95\% confidence level in the 
$m_A - \tan \beta$ plane.  Regions I and III correspond to those depicted 
in Fig.~1 with $m_{h^\pm}$ converted to $m_A$ taking into account the MSSM 
one-loop radiative corrections.  Region II represents a conservative 
estimate of the expected limit from LEP1 and LEP2 for $\tan \beta > 1$ 
(dotted line).  A slightly higher value of 62.5 GeV for $\tan \beta > 1$ 
has just recently been obtained by the ALEPH Collaboration\cite{15}.

\newpage
\begin{figure}[htb]
\centerline{ \DESepsf(amfs3.epsf width 15 cm) } \smallskip
\nonumber
\end{figure}

\end{document}